\documentclass[a4paper,11pt]{article}

\usepackage{amsmath}
\usepackage{amssymb}
\usepackage{color}
\usepackage{cite}
\usepackage{hyperref}
\usepackage{graphicx}
\usepackage{subfig}

\makeatletter
\@addtoreset{equation}{section}
\renewcommand{\theequation}{\thesection.\@arabic\c@equation}
\makeatother
\makeatletter
\renewcommand\appendix{\par
  \setcounter{section}{0}%
  \setcounter{subsection}{0}%
  \gdef\thesection{Appendix \@Alph\c@section }
  \renewcommand{\theequation}
  {\Alph{section}.\arabic{equation}}
}
\makeatother

\def \be {\begin{equation}}
\def \ee {\end{equation}}
\def \ba {\begin{array}}
\def \ea {\end{array}}
\def \bea{\begin{eqnarray}}
\def \eea{\end{eqnarray}}

\def \b {\beta}
\def \g {\gamma}

\def \d {\delta}
\def \D {\Delta}

\def \p {\partial}

\def \nn {\nonumber}

\def \hs {\hspace}

\def \inf {\infty}

\def \Tr {{\textrm{Tr}}}

\title{\textbf{ R\'enyi Entropy of Free Compact Boson on Torus }}
\author{
Bin Chen$^{1,2,3}$\footnote{bchen01@pku.edu.cn}\,
and
Jie-qiang Wu$^{1}$\footnote{jieqiangwu@pku.edu.cn}
}
\date{}

\begin{document}

\maketitle

\begin{center}
{\it
$^{1}$Department of Physics and State Key Laboratory of Nuclear Physics and Technology, Peking University, Beijing 100871, P.R.\! China\\
\vspace{2mm}
$^{2}$Collaborative Innovation Center of Quantum Matter, 5 Yiheyuan Rd, Beijing 100871, P.~R.~China\\
$^{3}$Center for High Energy Physics, Peking University, 5 Yiheyuan Rd, Beijing 100871, P.~R.~China
}
\vspace{10mm}
\end{center}

\begin{abstract}
In this paper, we reconsider the single interval R\'enyi entropy of a free compact scalar on a torus. In this case, the contribution to the entropy could be decomposed into classical part and quantum part. The classical part includes the contribution from all the saddle points,  while the quantum part is universal. After considering a different monodromy condition from the one in the literature, we re-evaluate the classical part of the R\'enyi entropy. Moreover, we  expand the entropy in the low temperature limit and find the leading thermal correction term which is consistent with the universal behavior suggested in \cite{Cardy:2014jwa}. Furthermore we investigate the large interval behavior of the entanglement entropy and show that the universal relation between the entanglement entropy and thermal entropy holds in this case.  
\end{abstract}

\baselineskip 18pt
\thispagestyle{empty}

\newpage

\section{Introduction}

Entanglement entropy is an important notion in many-body quantum system\cite{nielsen2010quantum,petz2008quantum}. For a bipartite system, the entanglement entropy of the subsystem $A$ is defined to be the Von Neumann entropy of the reduced density matrix
\be S_{EE}=-\Tr~\rho_A\log~\rho_A, \ee
where  the reduced density matrix $\rho_A=\Tr_B~\rho$ is obtained by smearing the degrees of freedom in $B$ complementary to A. When the total system is in the vacuum
$\rho=\mid 0\rangle\langle 0\mid$,  the entanglement entropy of $A$ equals to the one of its complement
\be S_{EE}(A)=S_{EE}(B). \ee
At a finite temperature, however, such
 equality does not hold anymore. The entanglement entropy has been  studied in various condensed matter systems\cite{Amico:2007ag}, and also in the context of AdS/CFT correspondence\cite{Ryu:2006bv,Ryu:2006ef}.  

The computation of the entanglement entropy from its definition becomes a formidable task when the number of degrees of freedom in the system is huge. Especially  in quantum field theory with infinite degrees of freedom, it is more convenient to compute the R\'enyi entropy via the Replica trick\cite{Callan:1994py}. The R\'enyi entropy is defined to be
\be S_n=-\frac{1}{n-1}\Tr \rho_A^n. \ee
It is related to the entanglement entropy by the relation
\be S_{EE}=\lim_{n\rightarrow 1} S_n \ee
if the analytic continuation of the limit is available. In field theory, the entropy is defined with respect to a spacial submanifold at a fixed time. In two dimensional spacetime, the submanifold could be interval(s). However, after Wick rotation, the Euclideanized field theory is defined in a complex plane and the R\'enyi entropy becomes
\be
S_n=-\frac{1}{n-1}\frac{Z_n}{Z_1^n},
\ee
in which  $Z_n$ is the partition function on a $n$-sheeted Riemann surface resulted from the pasting of $n$ complex plane along the branch cuts (intervals).

There has been a long history to study  the entanglement entropy in quantum field theory. In the dimension higher than two,  the entanglement entropy has been found to obey  the area law (for a nice review, see \cite{Eisert:2008ur}). But we are only allowed to compute the entanglement entropy analytically in very restricted situations, for example the one for sphere in free field theory.  In $(1+1)$ dimension, we can do better, especially for the field theory with conformal symmetry. For a 2D CFT on complex plane the R\'enyi entropy for one interval of length $\ell$ is universal and only depends on the central charge \cite{Calabrese:2004eu}
\be S_n=c\frac{n+1}{6n}\log \frac{l}{\epsilon}. \ee
 For more complicated cases, for example the multi-interval at zero temperature or single interval on a circle at finite temperature, the entanglement entropy and R\'enyi entropy depend on the details of the CFT. In \cite{Calabrese:2009ez,Calabrese:2010he},  the double interval R\'enyi entropies for free bosons and Ising model have been computed. 
 In \cite{Azeyanagi:2007bj,Herzog:2013py} the finite temperature R\'enyi entropy for free fermions has been discussed. In \cite{Datta:2013hba},  the finite temperature R\'enyi entropy for free bosons has been studied. Moreover, the treatment on the $W$-functions in the partition functions of free boson has been improved in \cite{Chen:2014hta}.

In this paper, we reconsider the R\'enyi entropy of free compact scalar on a torus. Our motivation is two-fold. First of all, in our study on the free noncompact scalar case\cite{Chen:2014hta}, we noticed at least two novel features, originated from the continuous spectrum of the theory. One is that the leading thermal correction at low temperature in this case takes a form different from the one suggested in \cite{Cardy:2014jwa}. This is because that the noncompact scalar has a degenerate vacuum, 
while the universal thermal correction found in \cite{Cardy:2014jwa} was based on  the assumption  that the CFT has a mass gap. 
The other one is that the universal relation between the large interval entanglement entropy and thermal entropy does not hold any more\cite{Chen:2014ehg,Chen:2014hta}. Since the noncompact free scalar could be taken as the large radius limit of the compact free scalar, it would be interesting to study the R\'enyi entropy of free compact scalar, which has a discrete spectrum. On the other hand, even though the R\'enyi entropy of free compact scalar has been computed in \cite{Datta:2013hba}, the detailed discussion on the low temperature or
large interval expansion has not been worked out. Moreover, we find that the classical part of the partition function actually depends on a different monodromy condition from the one in \cite{Datta:2013hba}. With the corrected classical partition function, we compute the R\'enyi entropy and do expansion in several limits, and rediscover the expected universal behaviors.

In the next section, we revaluate the R\'enyi entropy for free compact boson. We consider a slightly different monodromy condition to read the classical part of the partition function. Then in Section 3, we discuss a low temperature limit, and expand the R\'enyi entropy with respect to $e^{-2\pi \beta}$. We find the leading order is now consistent with the universal thermal correction suggested in \cite{Cardy:2014jwa}. In Section 4, we investigate  the small interval and large interval limits of the R\'enyi entropy,  and prove the universal relation  between entanglement and thermal entropies.

\section{Compact boson R\'enyi entropy}


For a free boson, the partition function on a Riemann surface can be decomposed into classical and quantum parts
\be\label{partition} Z=Z_{quantum}Z_{classical}. \ee
The classical part gets contributions from all the saddle points carrying different monodromy conditions
\be Z_{classical}=\sum_{\mathcal{M}} e^{-S_{E}({\mathcal{M}})}. \ee
 For quantum correction, we need to consider the perturbations around the classical saddle point and evaluate their contribution to the partition function. In general, for different classical solution their quantum corrections are different, but in the case of free bosons the quantum correction is universal so the partition function  can be decomposed into the classical and quantum part as in (\ref{partition}).

In this section, we compute the compact scalar partition functions on n-sheeted torus connected by a branch cut. The free complex scalar is compactified on a square torus of radius $R$. It obeys the boundary condition
\be\label{bound}
X(e^{2\pi i}z, e^{-2\pi i}\bar{z})=X(z,\bar{z})+2\pi R(m_1+im_2), \hs{3ex}m_1,m_2\in Z.
\ee
The quantum part of the partition function equals to \cite{Datta:2013hba,Chen:2014hta}
\be Z_{n,~quantum}=\frac{1}{\mid \eta(\tau)\mid^{4n}}\prod_{k=0}^{n-1}\frac{1}{\mid W_1^{1(k)}W_2^{2(k)}\mid}
\left(\frac{\vartheta_1^{'}(0\mid\tau)}{\vartheta_1(z_1-z_2\mid\tau)}\right)^{\frac{1}{6}n(1-\frac{1}{n^2})}
\left(\frac{\bar{\vartheta}_1^{'}(0\mid\bar{\tau})}{\bar{\vartheta}_1^{'}(\bar{z}_1-\bar{z}_2\mid\bar{\tau})}\right)
^{\frac{1}{6}n(1-\frac{1}{n^2})}, \ee
where we have already used the modular symmetry to simplify the expression. This result could be derived by using the Ward identity for the twist operators\cite{classical}.

For classical part, we need to find all of the classical solutions and calculate their action. For convenience we redefine the fields as
\be X^{(t,k)}(z,\bar{z})=\sum_{j=0}^{n-1}e^{\frac{2\pi i}{n}jk}X^{(j)}(z,\bar{z}), \ee
where $0\leq k<n$. For each redefined field $X^{(t,k)}(z,\bar{z})$, when the argument goes around the branch point $z_1$ or $z_2$, it gets an extra phase $e^{2\pi i\frac{k}{n}}$ or $e^{-2\pi i\frac{k}{n}}$. Moreover, the boundary condition (\ref{bound}) changes to
\be
X^{(t,k)}(e^{2\pi i}z,e^{-2\pi i}\bar{z})=e^{2\pi i\frac{k}{n}}X^{(t,k)}(z,\bar{z})+
v^{(k)},
\ee
where $v^{(k)}$ is a vector in the lattice $\Lambda_{\frac{k}{n}}$ defined by
\be
\Lambda_{\frac{k}{n}}\equiv \left\{2\pi R\sum_{j=0}^{n-1}e^{\frac{2\pi i}{n}jk}(m_{j,1}+im_{j,2}),\hs{3ex} m_{j,1},m_{j,2}\in Z\right\}.
\ee
These boundary conditions induce the following monodromy conditions
\bea &&\oint_{\gamma_a}dz \partial X^{(t,k)}(z)+\oint_{\gamma_a}d\bar{z} \bar{\partial}X^{(t,k)}=v_a^{(k)} \notag \\
&&\oint_{\gamma_a}dz \partial \bar{X}^{(t,k)}(z)+\oint_{\gamma_a}d\bar{z} \bar{\partial} \bar{X}^{(t,k)}=\bar{v}_a^{(k)}, \eea
where $\g_a$'s are the two cycles of worldsheet torus. In terms of cut differentials
 \bea &&\omega_1^{(k)}(z)=\vartheta_1(z-z_1\mid \tau)^{-(1-\frac{k}{n})}\vartheta_1(z-z_2\mid \tau)^{-\frac{k}{n}}
\vartheta_1(z-(1-\frac{k}{n})z_1-\frac{k}{n}z_2 \mid\tau) \notag \\
&&\omega_2^{(k)}(z)=\vartheta_1(z-z_1\mid\tau)^{-\frac{k}{n}}\vartheta_1(z-z_2\mid \tau)^{-(1-\frac{k}{n})}
\vartheta_1(z-\frac{k}{n}z_1-(1-\frac{k}{n})z_2 \mid \tau), \eea
the classical solutions can be written as
\bea &~&\partial X^{(t,k)}=a^{(k)}\omega_1^{(k)}(z),~~~
\bar{\partial}X^{(t,k)}=b^{(k)}\bar{\omega}_2^{(k)}(\bar{z}), \notag \\
&~&\partial \bar{X}^{(t,k)}=\tilde{a}^{(k)}\omega_2^{(k)}(z),
~~~\bar{\partial}\bar{X}^{(t,k)}=\tilde{b}^{(k)}\bar{\omega}_1^{(k)}(\bar{z}). \eea
Solving the monodromy condition, we get
\bea &&a^{(k)}=\frac{W_2^{2(k)}v_1-W_1^{2(k)}v_2}{\det W^{(k)}},~~~b^{(k)}=
\frac{-W_2^{1(k)}v_1+W_1^{1(k)}v_2}{\det W^{(k)}}, \notag \\
&&\tilde{a}^{(k)}=\frac{\bar{W}_1^{1(k)}\bar{v}_2-\bar{W}_2^{1(k)}\bar{v}_1}{\det \bar{W}^{(k)}}, ~~~
\tilde{b}^{(k)}=\frac{\bar{W}_2^{2(k)}\bar{v}_1-\bar{W}_1^{2(k)}\bar{v}_2}{\det \bar{W}^{(k)}}, \eea
where the $W$ functions are defined to be the integral of the cut differentials along different cycles. The definition and properties of the $W$ functions could be found in the Appendix.
With these results, the classical action for $X^{(t,k)}$ is just
\be\label{Sk} S^{(k)}=\frac{1}{4\pi n \alpha'}\frac{1}{\mid W_1^{1(k)} W_2^{2(k)}\mid}
(\mid v_1^{(k)}\mid^2 \mid W_2^{2(k)}\mid^2+\mid v_2^{(k)}\mid^2\mid W_1^{1(k)} \mid^2).
\ee

The above discussion is the same as the one in \cite{Datta:2013hba}. However, in  \cite{Datta:2013hba}, the classical action is further simplified using the relation between $W$ functions. The relation turns out to be problematic, as shown in \cite{Chen:2014hta}. Furthermore,  the lattice translation $v_1$ and $v_2$ were determined in \cite{Datta:2013hba} along the way in \cite{Calabrese:2009ez}. But notice that we are considering $n$-sheeted torus, which is very different from the $n$-sheeted Riemann surface got from two interval complex plane. Actually, the translation vectors are much simpler in the $n$-sheeted torus case.
  On this Riemann surface, the spacial cycles and time cycles on $n$ replicas build the canonical cycles, and all of the cycles on the Riemann surface can be generated by these cycles. Once we fix the monodromy around the canonical cycles, we can get all of the monodromy on the Riemann surface.

For the $n$-sheeted torus, the monodromy condition can be fixed as
\bea &&\Delta_1 X=2\pi R m_j \notag \\
&&\Delta_2 X=2\pi R n_j, \eea
where
\bea &&m_j=m_j^{(1)}+im_j^{(2)} \notag \\
 &&n_j=n_j^{(1)}+in_j^{(2)} \eea
are complex integers. If we transform into $X^{(t,k)}$ basis, then
\bea &&v_1^{(k)}=2\pi R\sum_{j=0}^{n-1}e^{2\pi ij\frac{k}{n}}m_j  \notag \\
&&v_2^{(k)}=2\pi R\sum_{j=0}^{n-1}e^{2\pi ij\frac{k}{n}}n_j. \eea
Taking these relations into (\ref{Sk}) and summing over all the twist fields, we get
\bea S_{cl}&=&\sum_{r=1,2}\frac{\pi R^2}{ n\alpha'}(\sum_k\left| \frac{W_2^{2(k)}}{W_1^{1(k)}}\right|\sum_{j=0}^{n-1}\sum_{j^{'}=0}^{n-1}
\cos 2\pi(j-j^{'})\frac{k}{n}\cdot m_j^{(r)}m_{j^{'}}^{(r)} \notag \\
&~&+\sum_k\left| \frac{W_1^{1(k)}}{W_2^{2(k)}}\right|\sum_{j=0}^{n-1}\sum_{j^{'}=0}^{n-1}
\cos 2\pi(j-j^{'})\frac{k}{n}\cdot n_j^{(r)}n_{j^{'}}^{(r)}).
\eea
For convenience, let us define
\bea &&A_{jj^{'}}=\sum_{k=0}^{n-1}\left|\frac{W_2^{2(k)}}{W_1^{1(k)}}\right| \cos2\pi(j-j^{'})\frac{k}{n} \notag \\
&&B_{jj^{'}}=\sum_{k=0}^{n-1}\left|\frac{W_1^{1(k)}}{W_2^{2(k)}}\right| \cos2\pi (j-j^{'})\frac{k}{n}. \eea
We can diagnose the two matrices with the same matrix
\be U_{jk}=e^{2\pi ij\frac{k}{n}}, \ee
and they have different eigenvalues
\bea &&U^{-1}\cdot A\cdot U=diag(n\big|  \frac{W_2^{2(0)}}{W_1^{1(0)}}\big| ,n\big| \frac{W_2^{2(1)}}{W_1^{1(1)}}\big|,...
n\big| \frac{W_2^{2(n-1)}}{W_1^{1(n-1)}}\big|) \notag \\
&&U^{-1}\cdot B\cdot U=diag(n\big|  \frac{W_1^{1(0)}}{W_2^{2(0)}}\big| ,n\big| \frac{W_1^{1(1)}}{W_2^{2(1)}}\big|,...
n\big| \frac{W_1^{1(n-1)}}{W_2^{2(n-1)}}\big|), \eea
so
\be B^{-1}=\frac{1}{n^2}A. \ee

We can calculate the classical contribution of the partition function
\bea Z_{classical}&=&\sum_{m_j^{(r)},n_j^{(r)}}\exp[-\frac{\pi R^2}{\alpha'n}\sum_{r=1,2}
\sum_{k=0}^{n-1}\left| \frac{W_2^{2(k)}}{W_1^{1(k)}}\right|
\sum_{j=0}^{n-1}\sum_{j^{'}=0}^{n-1}
\cos2\pi (j-j^{'})\frac{k}{n} \cdot m_j^{(r)}m_{j^{'}}^{(r)}
\notag \\
&~&-\frac{\pi R^2}{\alpha'n}\sum_{r=1,2}
\sum_{k=0}^{n-1}\left| \frac{W_1^{1(k)}}{W_2^{2(k)}}\right|
\sum_{j=0}^{n-1}\sum_{j^{'}=0}^{n-1}
\cos2\pi (j-j^{'})\frac{k}{n} \cdot
 n_j^{(r)}n_{j^{'}}^{(r)}]
 \notag \\
&=&\big(\sum_{m_j,n_j}\exp[-\frac{\pi R^2}{\alpha'n}
\sum_{k=0}^{n-1}\left| \frac{W_2^{2(k)}}{W_1^{1(k)}}\right|
\sum_{j=0}^{n-1}\sum_{j^{'}=0}^{n-1}
\cos2\pi (j-j^{'})\frac{k}{n} \cdot m_jm_{j^{'}} \notag \\
&~&-\frac{\pi R^2}{\alpha'n}
\sum_{k=0}^{n-1}\left| \frac{W_1^{1(k)}}{W_2^{2(k)}}\right|
\sum_{j=0}^{n-1}\sum_{j^{'}=0}^{n-1}
\cos2\pi (j-j^{'})\frac{k}{n} \cdot
 n_jn_{j^{'}}] \big)^2
\notag \\
&=&\frac{\alpha^{'n}}{R^{2n}}\prod_{k=0}^{n-1}\left| \frac{W_2^{2(k)}}{W_1^{1(k)}} \right| \left(\sum_{m_j,p_j}\exp\left\{-\frac{\pi R^2}{\alpha'n}
\sum_{k=0}^{n-1}\left| \frac{W_2^{2(k)}}{W_1^{1(k)}}\right|
\sum_{j=0}^{n-1}\sum_{j^{'}=0}^{n-1}
\cos2\pi (j-j^{'})\frac{k}{n} \cdot m_jm_{j^{'}} \right.\right.\notag \\
&~&\left.\left. -\frac{\alpha'\pi}{R^2n}
\sum_{k=0}^{n-1}\left|\frac{W_2^{2(k)}}{W_1^{1(k)}} \right|
\sum_{j=0}^{n-1}\sum_{j^{'}=0}^{n-1}
\cos2\pi(j-j^{'})\frac{k}{n}\cdot p_jp_{j^{'}}\right\}\right)^2.
\eea
For the last equation, we have used higher dimensional Poisson re-summation. It is remarkable that the classical partition function depends explicitly on the $W$ functions and then on the interval. This is very different from the result in \cite{Datta:2013hba}.  Combined with the quantum part, the full partition function reads
\bea\label{Zn} Z_n&=&Z_{classical}Z_{quantum} \notag \\
&=&c_n\frac{1}{\mid\eta(\tau)\mid^{4n}}\prod_{k=0}^{n-1}\frac{1}{\mid W_1^{1(k)}\mid^2}
\big( \frac{\vartheta_1^{'}(0\mid\tau)}{\vartheta_1(z_1-z_2\mid\tau)} \big)^{\frac{1}{6}n(1-\frac{1}{n^2})}
\big( \frac{\bar{\vartheta}_1^{'}(0\mid\bar{\tau})}
{\bar{\vartheta}_1(\bar{z}_1-\bar{z}_2\mid\bar{\tau})} \big)^{\frac{1}{6}n(1-\frac{1}{n^2})} \notag \\
&~&\cdot \left(\sum_{m_j,p_j}\exp\left\{-\frac{\pi R^2}{\alpha'n}
\sum_{k=0}^{n-1}\left| \frac{W_2^{2(k)}}{W_1^{1(k)}}\right|
\sum_{j=0}^{n-1}\sum_{j^{'}=0}^{n-1}
\cos2\pi (j-j^{'})\frac{k}{n} \cdot m_jm_{j^{'}} \right.\right.\notag \\
&~&\left.\left. -\frac{\alpha'\pi}{R^2n}
\sum_{k=0}^{n-1}\left|\frac{W_2^{2(k)}}{W_1^{1(k)}} \right|
\sum_{j=0}^{n-1}\sum_{j^{'}=0}^{n-1}
\cos2\pi(j-j^{'})\frac{k}{n}\cdot p_jp_{j^{'}}\right\}\right)^2,
\eea
where  other coefficients have been absorbed  into $c_n$. This is the main result of this paper.

\section{Low temperature limit}

For the single interval R\'enyi entropy at finite temperature, there is an universal thermal correction coming from the lowest excitation\cite{Cardy:2014jwa}
\be
\d S_n=\frac{gn}{1-n}\left(\frac{\sin\frac{\pi l}{L}}{n\sin\frac{\pi l}{nL}}\right)^{2\D}e^{-2\pi \D\b/L}+\cdots
\ee
where $\D$ is the  scaling dimension of the excitations and $g$ is their degeneracy. This relation has been checked to be true for the vacuum module in the context of AdS$_3$/CFT$_2$ correspondence\cite{small}. However, it has shown to break down for the noncompact free scalar, which has a continuous spectrum.
In this section, we study the low temperature limit of the partition function and R\'enyi entropy of compact free scalar, and check this relation.

For simplicity, we assume $\frac{\alpha'}{R^2}<1$, that means the momentum mode has lower dimension than the descendant of vacuum module. We expand the results with respect to $q=e^{-2\pi \beta}$,
\bea \eta(\tau)&=&q^{\frac{1}{24}}(1+O(q)) \\
 \vartheta_1^{'}(0)&=&2\pi q^{\frac{1}{8}}(1+O(q)) \\
 \vartheta_1(z_1-z_2\mid\tau)&=&2q^{\frac{1}{8}}\sin\pi(z_1-z_2)(1+O(q)) \eea
\bea \frac{W_2^{2(k)}}{W_1^{1(k)}}=i\beta+\int_{z_1}^{z_2}dt
\frac{2i ( \sin\frac{\pi k}{n}(t-z_1)  ) (\sin\pi(1-\frac{k}{n})(t-z_1))}{\sin\pi(t-z_1)}+O(q) \eea
For the summation in (\ref{Zn}), the leading and next leading contribution come from $m_j=p_j=0$ and $m_j=0,p_i=\pm 1,p_j=0~\mbox{for}~j\neq i$. When all of $m_j$ and $p_j$ are zero, then the exponential is 1. When only $p_i=\pm1$, the terms on the exponential contribute
\bea &~&-\frac{\alpha'\pi}{R^2n}\sum_k ^{n-1}\left| \frac{W_2^{2(k)}}{W_1^{1(k)}}\right| \notag \\
&=&-\frac{\alpha'\pi}{R^2n}\sum_k ^{n-1}
(\beta+\int_{z_1}^{z_2}dt
\frac{2\sin\frac{\pi k}{n}(t-z_1)\sin\pi(1-\frac{k}{n})(t-z_1)}{\sin\pi(t-z_1)})+O(q) \notag \\
&=&-\frac{\alpha'\pi}{R^2}\beta -\frac{\alpha'}{R^2}\log~
\frac{n\sin\frac{\pi}{n}(z_2-z_1)}{\sin\pi(z_2-z_1)}+O(q).
\eea
Thus the leading and subleading contributions in the summation give
\be 1+2ne^{-\frac{\pi \alpha'\beta}{R^2}}(\frac{\sin\pi(z_2-z_1)}{n\sin\frac{\pi}{n}(z_2-z_1)})^
{\frac{\alpha'}{R^2}}+O(e^{ \frac{2\pi \alpha'\beta}{R^2}})\ee
and the partition function is approximately
\be Z_n=c_n\frac{1}{q^{\frac{n}{6}}}(\frac{\pi}{\sin\pi l})^{\frac{1}{3}n(1-\frac{1}{n^2})}
(1+4ne^{-\frac{\pi\alpha'\beta}{R^2}}
(\frac{\sin\pi l}{n\sin\frac{\pi}{n}l})^{\frac{\alpha'}{R^2}}+
O(e^{ \frac{2\pi \alpha'\beta}{R^2}})) \ee
and
\be S_n=c_n+\frac{n+1}{3n}-\frac{1}{n-1}4n
(\frac{\sin\pi l}{n\sin\frac{\pi}{n}l})^{\frac{\alpha'}R^2}
e^{-\frac{\pi \alpha'\beta}{R^2}}
+O(e^{-\frac{2\pi \alpha'\beta}{R^2}}).
\ee
This result is in consistent with the universal behavior suggested in \cite{Cardy:2014jwa}. For the complex scalar, the central charge is 2. There are four degeneracies for the lowest excitation  states, with the conformal dimension
$(\frac{\alpha'}{4R^2},\frac{ \alpha'}{4R^2})$.

\section{Relation between thermal entropy and entanglement entropy}

In \cite{Azeyanagi:2007bj}, it was suggested that when the interval becomes very large, the entanglement entropy and thermal entropy could be related by
\be\label{thermal} S_{th}=\lim_{\epsilon\to 0}(S_{EE}(1-\epsilon)-S_{EE}(\epsilon)). \ee
In \cite{Chen:2014ehg,Chen:2014hta}, this relation has been proved for a general CFT with discrete spectrum. For the free compact scalar case at hand, it has discrete spectrum so should satisfy the relation. We can prove it directly by expanding the $W$ functions.

Since it is not obvious on how to take $n\to 1$ limit on the R\'enyi entropy, we do not have explicit form of the entanglement entropy. Instead,  we first try to study the $\epsilon \to 1$ limit  and then take the $n\to 1$ limit. We assume that taking the limits of $n\rightarrow 1$ and $\epsilon\rightarrow 1$ is commutable.

For small interval, we have
\be W_1^{1(k)}=1+O(z_1-z_2), \hs{3ex}  W_2^{2(k)}=i\beta+O(z_1-z_2), \ee
and thus
\bea Z_n&=&c_nl^{-\frac{1}{3}n(1-\frac{1}{n^2})}\frac{1}{|\eta(\tau)|^{4n}}
\cdot(\sum_{m_j,p_j}\exp [-\frac{\pi R^2}{\alpha'}m_j^2
-\frac{\alpha'\pi}{R^2}p_j^2]+O(z_1-z_2))^2 \notag \\
&=&c_nl^{-\frac{1}{3}n(1-\frac{1}{n^2})}
[(\frac{1}{|\eta(\tau)|^2}\sum_{m,p} \exp[-\frac{\pi R^2}{\alpha'}m^2
-\frac{\alpha'\pi}{R^2}p^2])^{2n} +O(l)] \notag \\
&=&c_nl^{-\frac{1}{3}n(1-\frac{1}{n^2})}[Z_1^n+O(l)]
\eea

For large interval, we first analyze the summation terms in the exponential. We only extract the leading terms with respect to $z_1-z_2$. Since the eigenvalues for the matrix $A$ are $n|\frac{W_2^{2(k)}}{W_1^{1(k)}}|, k=0,\cdots n-1$,  in the limit $z_1\rightarrow z_2$, all the terms are suppressed but the terms with $m_1=m_2=...m_n=m, p_1=p_2=...p_n=p$. This fact is consistent with the observation in \cite{Chen:2014ehg,Chen:2014hta} that the excitations on different replicas should be the same in order to give nonvanishing contributions in the large interval limit. Therefore the summation yields
\be \sum_{m,p} \exp[-\frac{\pi R^2n\beta}{\alpha'}m^2-\frac{\alpha'n\beta}{R^2}p^2]. \ee
In this case, when $z_1\rightarrow z_2$, the prefactor before the exponential  goes to
\bea &~&c_nl^{-\frac{1}{3}n(1-\frac{1}{n^2})}
\frac{1}{|\eta(\tau)|^{4n}}\prod_{k=0}^{n-1}|\frac{2\sin\frac{\pi k}{n}\eta(\tau)^3}
{\vartheta(-\frac{k}{n}|\tau)}|^2 \notag \\
\eea
This factor could be simplified more
\bea &~&\frac{1}{|\eta(i\beta)|^{4n}}\prod_{k=1}^{n-1}|\frac{2\sin\frac{\pi k}{n}\eta(i\beta)^3}
{\vartheta_1(-\frac{k}{n}|i\beta)}|^2 \notag \\
&=&\prod_{k=1}^{n-1}(2\sin\frac{\pi k}{n})^2(\beta^{-\frac{1}{2}}\eta(\frac{i}{\beta}))^{2n-6}
 \prod_{k=1}^{n-1}\left| \frac{1}{\beta^{-\frac{1}{2}}e^{-\frac{k^2}{n^2\beta}}
\vartheta_1(-\frac{k}{i\beta n}|\frac{i}{\beta})}\right|^2 \notag \\
&=&\prod_{k=1}^{n-1}(2\sin\frac{\pi k}{n})^2\beta^2 \frac{1}{\eta(\frac{i}{n\beta})^4} \notag \\
&=&\frac{1}{n^2}\prod_{k=1}^{n-1}(2\sin\frac{\pi k}{n})^2\frac{1}{\eta(in\beta)^4}
\eea
As
\be \frac{1}{n^2}\prod_{k=1}^{n-1}(2\sin\frac{\pi k}{n})^2=1,  \ee
we find that  for the large interval
\be Z_n=c_n l^{\frac{1}{3}n(1-\frac{1}{n^2})}(Z_1[n\beta]+O(l^{\lambda})), \ee
where $\lambda<1$ and $O(l^{\lambda})$ terms coming from the contributions of the primary fields  in the operator product expansion of two twist operators.

Now we find the similar results as the ones in \cite{Chen:2014ehg,Chen:2014hta}. It is easy and straightforward to  prove the relation (\ref{thermal}) between the thermal entropy and the entanglement entropy. Actually,
\bea
\lefteqn{\lim_{l\to 0}(S_{EE}(1-l)-S_{EE}(l))} \nn \\
&=&-\lim_{n\rightarrow1}\frac{1}{n-1}(\log Z_1[n\b]-n \log Z_1[\b])\nn\\
&=&\log Z[\b]-\frac{1}{\b}\frac{Z'[\b]}{Z[\b]}\nn\\
&=& S_{th}. \nn
\eea



\vspace*{10mm}
\noindent {\large{\bf Acknowledgments}}\\

 The work was in part supported by NSFC Grant No.~11275010, No.~11335012 and No.~11325522.
\vspace*{5mm}

\begin{appendix}

\section{$W$ functions}
In this Appendix, we list some useful results for the $W$ functions and study their properties in various limits. The $W$ functions are defined as the line integral of the cut differentials
\bea &&W_1^{1(k)}=\int_0^1dz\vartheta_1(z-z_1\mid\tau)^{-(1-\frac{k}{n})}
\vartheta_1(z-z_2\mid\tau)^{-\frac{k}{n}}
\vartheta_1(z-(1-\frac{k}{n})z_1-\frac{k}{n}z_2\mid\tau) \notag \\
&&W_1^{2(k)}=\int_0^1d\bar{z}\bar{\vartheta}_1(\bar{z}-\bar{z}_1\mid\tau)^{-\frac{k}{n}}
\bar{\vartheta}_1(\bar{z}-\bar{z}_2\mid\tau)^{-(1-\frac{k}{n})}
\bar{\vartheta}_1(\bar{z}-\frac{k}{n}\bar{z}_1-(1-\frac{k}{n})\bar{z}_2\mid\tau)
\notag \\
&&W_2^{1(k)}=\int_0^{\tau}dz \vartheta_1(z-z_1\mid\tau)^{-(1-\frac{k}{n})}
\vartheta_1(z-z_2\mid\tau)^{-\frac{k}{n}}
\vartheta_1(z-(1-\frac{k}{n})z_1-\frac{k}{n}z_2\mid\tau) \notag \\
&&W_2^{2(k)}=\int_0^{\bar{\tau}}d\bar{z}\bar{\vartheta}_1(\bar{z}-\bar{z}_1\mid\tau)^{-\frac{k}{n}}
\bar{\vartheta}_1(\bar{z}-\bar{z}_2\mid\tau)^{-(1-\frac{k}{n})}
\bar{\vartheta}_1(\bar{z}-\frac{k}{n}\bar{z}_1-(1-\frac{k}{n})\bar{z}_2\mid\tau).
\eea
As we have chosen the modular parameter to be pure imaginary $\tau = i\b$, the $W$ functions are related by
\be W_1^{1*}=W_1^1=W_1^2,~~~W_2^{1*}=-W_2^1=W_2^2. \ee

\begin{figure}[tbp]
\centering
\subfloat[A cycle]{\includegraphics[width=5cm]{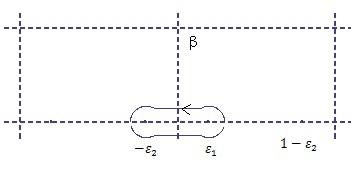}\label{Acycle}}
\quad
\subfloat[B cycle]{\includegraphics[width=5cm]{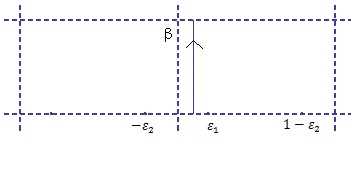}\label{Bcycle}}
\caption{There are two contour integral paths.  }
\end{figure}

\subsection{Low temperature expansion}

The low temperature expansion for $W_1^{1(k)}$ is
\bea &~&\lefteqn{W_1^{1(k)}} \notag \\
&=&(\int_0^{z_1}+\int_{z_1}^{z_2}+\int_{z_2}^{1})du\vartheta_1(z-z_1\mid\tau)^{-(1-\frac{k}{n})}
\vartheta_1(z-z_2\mid\tau)^{-\frac{k}{n}}
\vartheta_1(z-(1-\frac{k}{n})z_1-\frac{k}{n}z_2\mid\tau) \notag \\
&=&\int_{z_2-1}^{z_1}dz\vartheta_1(z-z_1\mid\tau)^{-(1-\frac{k}{n})}
(\vartheta_1(z-(z_2-1)\mid\tau)e^{\pi i})^{-\frac{k}{n}}
\vartheta_1(z-(1-\frac{k}{n})z_1-\frac{k}{n}z_2)\mid\tau) \notag \\
&=&-\frac{e^{-\frac{\pi ik}{n}}}{1-e^{-\frac{2\pi ik}{n}}}\oint_A dz
(\sin\pi(z-z_1))^{-(1-\frac{k}{n})}(\sin\pi(z-(z_2-1)))^{-\frac{k}{n}}
\sin\pi(z-(1-\frac{k}{n})z_1-\frac{k}{n}z_2) +O(q)\notag \\
&=&\frac{e^{-\pi i\frac{k}{n}}}{-1+e^{-2\pi i\frac{k}{n}}}(\oint_{\infty} \frac{du}{2\pi iu} e^{-\frac{k}{n}\pi i}
(u-u_1)^{-(1-\frac{k}{n})}(u-u_2)^{-\frac{k}{n}}(u-u_1^{(1-\frac{k}{n})}u_2^{\frac{k}{n}}) \notag \\
&~&-\oint_{0} \frac{du}{2\pi iu} e^{-\frac{k}{n}\pi i}
(u-u_1)^{-(1-\frac{k}{n})}(u-u_2)^{-\frac{k}{n}}(u-u_1^{(1-\frac{k}{n})}u_2^{\frac{k}{n}})
+O(q)\notag \\
&=&\frac{e^{-\pi i\frac{k}{n}}}{-1+e^{-2\pi i\frac{k}{n}}}(\oint_{\infty} e^{-\frac{k}{n}\pi i}
(1-(e^{\pi i}e^{2\pi iz_1})^{-(1-\frac{k}{n})}(e^{\pi i}e^{2\pi i(z_2-1)})^{-\frac{k}{n}})+O(q) \notag \\
&=&1+O(q).
\eea
In the forth equation we have used a conformal transformation
\be u=e^{2\pi iz}, \ee
and in the fifth equation we have changed the integral contour to the one surrounding the infinity and the origin. For $W_2^{2(k)}$, we have the expansion
\bea &~&\bar{W}_2^{2(k)} \notag \\
&=&\int_0^{i\beta} dz
\vartheta(z-z_1\mid\tau)^{-\frac{k}{n}}\vartheta(z-z_2\mid\tau)^{-(1-\frac{k}{n})}
\vartheta(z-(\frac{k}{n}z_1+(1-\frac{k}{n})z_2)\mid\tau) \notag \\
&=&\int_0^{i\beta} dz
((1-e^{2\pi i(z-z_1)})(1-qe^{-2\pi i(z-z_1)}))^{-\frac{k}{n}}
((1-e^{2\pi i(z-z_2)})(1-qe^{-2\pi i(z-z_2)}))^{-(1-\frac{k}{n})} \notag \\
&~&((1-e^{2\pi i(z-(\frac{k}{n}z_1+(1-\frac{k}{n})z_2))})(1-qe^{-2\pi i(z-(\frac{k}{n}z_1+(1-\frac{k}{n})z_2))}))
+O(q).
 \eea
We can analytically expand the equations with respect to $e^{2\pi iz}$ and the summations in  the expansions still converge. There appear several kinds of terms in the expansion
\be 1,~~~e^{2\pi m iz},~~~q^{m}e^{-2\pi miz},~~~q^{m}e^{2\pi i(-m+n)z}. \ee
After being integrated, the first term gives $i\beta$. The second and third terms give
\be \int_0^{i\beta} e^{2\pi miz}=\frac{1}{2\pi mi}e^{2\pi miz}\mid_0^{i\beta}=\frac{1}{2\pi mi}(q^m-1), \ee
\be \int_0^{i\beta} q^{m}e^{-2\pi miz}=\frac{1}{-2\pi mi}q^{m}e^{-2\pi miz}\mid_0^{i\beta}
=\frac{1}{-2\pi mi}q^{m}(q^{-m}-1),\ee
\be \int_0^{i\beta} q^{m}e^{2\pi i(-m+n)z}=O(q). \ee
If we are interested in the leading contribution with respect to $q$, we only need to consider the first three kinds of terms. In the end, we have
\bea &~&\bar{W}_2^{2(k)} \notag \\
&=&i\beta+\int_0^{i\beta}dz[(1-e^{2\pi i(z-z_1)})^{-\frac{k}{n}}(1-e^{2\pi i(z-z_2)})^{-(1-\frac{k}{n})}
(1-e^{2\pi i(z-\frac{k}{n}z_1-(1-\frac{k}{n})z_2)})-1] \notag \\
&~&+\int_0^{i\beta}dz[(1-qe^{-2\pi i(z-z_1)})^{-\frac{k}{n}}(1-qe^{-2\pi i(z-z_2)})^{-(1-\frac{k}{n})}
(1-e^{-2\pi i(z-\frac{k}{n}z_1-(1-\frac{k}{n})z_2)})-1] +O(q) \notag \\
&=&i\beta+\int_{-i\inf}^{i\inf}dz [(\sin\pi(z-z_1))^{-\frac{k}{n}}(\sin\pi(z-z_2))^{-(1-\frac{k}{n})}
(\sin\pi(z-\frac{k}{n}z_1-(1-\frac{k}{n})z_2)) -1]+O(q) \notag \\
\eea
To deal with the integral in the last relation, we define
\be F(z_1,z_2)=\int_{-i\inf}^{i\inf}dz [(\sin\pi(z-z_1))^{-\frac{k}{n}}(\sin\pi(z-z_2))^{-(1-\frac{k}{n})}
(\sin\pi(z-\frac{k}{n}z_1-(1-\frac{k}{n})z_2)) -1], \ee
which only depends on $z_1-z_2$.
Considering
\be \frac{\partial_{z_1}F(z_1,z_2)}{\sin\pi(1-\frac{k}{n})(z_1-z_2)}
=\pi\frac{k}{n}\int_{-i\inf}^{i\inf}(\sin\pi(z-z_1))^{-\frac{k}{n}-1}(\sin\pi(z-z_2))^{-(1-\frac{k}{n})} \ee
\be \partial_{z_2}\frac{\partial_{z_1}F(z_1,z_2)}{\sin\pi(1-\frac{k}{n})(z_1-z_2)}
=\pi^2\frac{k}{n}(1-\frac{k}{n})\int_{-i\inf}^{i\inf}
(\sin\pi(z-z_1))^{-\frac{k}{n}-1}(\sin\pi(z-z_2))^{-(2-\frac{k}{n})}\cos\pi(z-z_2), \ee
we find
\bea \partial_{z_2}F&=&-\frac{1-\frac{k}{n}}{\frac{k}{n}}
\sin\frac{\pi k}{n}(z_2-z_1)\cos\pi(z_2-z_1)\frac{\partial_{z_2}F}{\sin\pi(1-\frac{k}{n})(z_1-z_2)} \notag \\
&~&-\frac{1}{\pi \frac{k}{n}}\sin\frac{\pi k}{n}(z_2-z_1)\sin\pi(z_2-z_1)
\partial_{z_2}(\frac{\partial_{z_2}F}{\sin\pi(1-\frac{k}{n})(z_1-z_2)})
\eea
Define
\be T=\partial_{z_2}F(z_1,z_2), \ee
we have the equation
\be \frac{\partial_{z_2}T}{T}=\pi \frac{k}{n}\frac{\cos\pi\frac{k}{n}(z_2-z_1)}{\sin\pi\frac{k}{n}}
+\pi\frac{n-k}{n}\frac{\cos\pi(1-\frac{k}{n})(z_2-z_1)}{\sin\pi(1-\frac{k}{n})(z_2-z_1)}
-\pi\frac{\cos\pi(z_2-z_1)}{\sin\pi(z_2-z_1)}, \ee
which has the solution
\be T=C\cdot\frac{\sin\frac{\pi k}{n}(z_2-z_1)\sin\pi(1-\frac{k}{n})(z_2-z_1)}{\sin\pi(z_2-z_1)}. \ee
Comparing with the direct evaluation of $\p_{z_2}F(z_1,z_2)$ for $z_2\rightarrow z_1$, we have
\be C=2i. \ee
Therefore,
\be F(z_1,z_2)=\int_{z_1}^{z_2}dt
\frac{2i\sin\frac{\pi k}{n}(t-z_1)\sin\pi(1-\frac{k}{n})(t-z_1)}{\sin\pi(t-z_1)}, \ee
and
\be\label{W} \bar{W}_2^{2(k)}=i\beta+\int_{z_1}^{z_2}dt
\frac{2i\sin\frac{\pi k}{n}(t-z_1)\sin\pi(1-\frac{k}{n})(t-z_1)}{\sin\pi(t-z_1)}+O(q). \ee

\subsection{Small and large interval limits}

To study the relation between thermal entropy and entanglement entropy, we need to study the small interval limit and large interval limit of the $W$ functions.
For small interval, it is simple to calculate. When $z_1=z_2$, the integrand is 1, so
\be W_1^{1(k)}=1~~~\bar{W}_2^{2(k)}=i\beta. \ee
For large interval, we set $z_1\rightarrow 0$, $z_2\rightarrow 1$,
\bea W_1^{1(k)}&=&
-\frac{e^{-\frac{\pi ik}{n}}}{1-e^{-\frac{2\pi ik}{n}}}\oint_A dz
\vartheta_1(z-z_1\mid\tau)^{-(1-\frac{k}{n})}
\vartheta_1(z-(z_2-1)\mid\tau)^{-\frac{k}{n}}
\vartheta_1(z-(1-\frac{k}{n})z_1-\frac{k}{n}z_2\mid\tau) \notag \\
&=&-\frac{e^{-\frac{\pi ik}{n}}}{1-e^{-\frac{2\pi ik}{n}}}\oint_A dz
\vartheta_1(z)^{-1}\vartheta_1(z-\frac{k}{n}\mid\tau)+O(z_1-(z_2-1)) \notag \\
&=&-\frac{1}{2\sin\pi\frac{k}{n}}\eta(\tau)^{-3}\vartheta_1(-\frac{k}{n}\mid\tau)+O(z_1-(z_2-1)).
\eea
Let us evaluate the most singular term in $W_2^{2(k)}$ when $z_1\rightarrow 0$ and $z_2\rightarrow 1$
\bea \bar{W}_2^{2(k)}&=&e^{-(1-\frac{k}{n})\pi i}
\int_0^{i\beta}dz \vartheta_1(z-z_1|\tau)^{-\frac{k}{n}} \vartheta_1(z-(z_2-1)|\tau)^{-(1-\frac{k}{n})}
\vartheta_1(z-\frac{k}{n}z_1-(1-\frac{k}{n})z_2|\tau). \notag
\eea
There is a $(z-z_1)$ term in $\vartheta_1(z-z_1|\tau)$, and a $(z-(z_2-1))$ term in $\vartheta(z-(z_2-1)|\tau)$. The most singular terms in the integral come from the integral range  near the origin. In this limit
\bea \vartheta_1(z-z_1|\tau)&\sim & 2\pi(z-z_1)\eta(\tau) \\
 \vartheta_1(z-(z_2-1)|\tau)&\sim & 2\pi(z-(z_2-1))\eta(\tau) \\
\vartheta_1(z-\frac{k}{n}z_1-(1-\frac{k}{n})z_2|\tau) &\sim & \vartheta_1(-(1-\frac{k}{n})|\tau), \eea
then
\bea \bar{W}_2^{2(k)}\sim e^{-(1-\frac{k}{n})\pi i}
\int_{-iM}^{iM}dz \frac{1}{2\pi}(z-z_1)^{-\frac{k}{n}}(z-(z_2-1))^{-(1-\frac{k}{n})}\eta(\tau)^{-3}
\vartheta_1(-(1-\frac{k}{n})\mid\tau), \eea
where $(z_2-z_1)\ll M\ll \beta$. Since
\be \int_{-iM}^{iM}dz(z-z_1)^{-\frac{k}{n}}(z-(z_2-1))^{-(1-\frac{k}{n})}\sim
-(1-e^{-2\pi i\frac{k}{n}})\log~(z_1-(z_2-1)), \ee
 the leading singular term in $W_2^{2(k)}$ is
\be \bar{W}_2^{2(k)}\sim i\frac{\sin\frac{\pi k}{n}}{\pi}\eta(\tau)^{-3}\vartheta_1(-(1-\frac{k}{n})|\tau)
\log~(z_1-(z_2-1)), \ee
and
\be \left|\frac{W_2^{2(k)}}{W_1^{1(k)}} \right|\sim \frac{2\sin\frac{\pi k}{n}}{\pi} (-\log(z_1-(z_2-1))). \ee

\end{appendix}



\end{document}